\documentclass[11pt]{article} 
\pdfoutput=1

\usepackage[utf8]{inputenc}
\usepackage[T1]{fontenc} 
\usepackage{lmodern}
\usepackage[english]{babel}           
\usepackage{cancel}
\usepackage[pdftex]{graphicx}
\usepackage{epsfig}
\usepackage{graphicx}
\usepackage{comment}
\usepackage{latexsym}
\usepackage{hyperref}
\usepackage{amsmath}
\usepackage{mathrsfs}
\usepackage[usenames, dvipsnames]{color}
\usepackage{amsbsy}
\usepackage{amssymb}
\usepackage{amsthm}
\usepackage{amsfonts}
\usepackage{cite}
\usepackage{enumitem}
\usepackage{xcolor}
\usepackage{color}
\usepackage{mathtools}
\usepackage{caption} 
\usepackage{subcaption} 
\usepackage{euscript} 
\usepackage{float}
\usepackage{diagbox}
\usepackage[title]{appendix}
\usepackage{tabularx}

\renewcommand\d{{\rm d}}

\renewcommand{\S}{{\cal S}}

\newcommand{\be}{\begin{equation}}
\newcommand{\ee}{\end{equation}}

\renewcommand{\thefootnote}{\fnsymbol{footnote}}

\newcommand{\Eq}[1]{Eq.~\eqref{#1}}
\newcommand{\Eqs}[1]{Eqs.~\eqref{#1}}

\newcommand{\Refs}[1]{Refs.~\cite{#1}}
\newcommand{\Sect}[1]{Sect.~\ref{#1}}

\newcommand{\ie}{{\em i.e.} }
\newcommand{\eg}{{\em e.g.} }
\newcommand{\via}{{\it via} }

\newcommand{\where}{\mbox{where}}

\renewcommand{\and}{\mbox{and}}

\newcommand{\bm}{\boldmath} 

\newcommand{\hi}{{\hat \imath}}

\catcode`\@=11
\def\marginnote#1{}
\newcount\hour
\newcount\minute
\newtoks\amorpm
\hour=\time\divide\hour by60 \minute=\time{\multiply\hour by60
\global\advance\minute by-\hour}
\edef\standardtime{{\ifnum\hour<12 \global\amorpm={am}%
        \else\global\amorpm={pm}\advance\hour by-12 \fi
        \ifnum\hour=0 \hour=12 \fi
        \number\hour:\ifnum\minute<10 0\fi\number\minute\the\amorpm}}
\edef\militarytime{\number\hour:\ifnum\minute<10 0\fi\number\minute}
\def\draftlabel#1{{\@bsphack\if@filesw {\let\thepage\relax
   \xdef\@gtempa{\write\@auxout{\string
      \newlabel{#1}{{\@currentlabel}{\thepage}}}}}\@gtempa
   \if@nobreak \ifvmode\nobreak\fi\fi\fi\@esphack}
        \gdef\@eqnlabel{#1}}
\def\@eqnlabel{}
\def\@vacuum{}
\def\draftmarginnote#1{\marginpar{\raggedright\scriptsize\tt#1}}
\def\draft{\oddsidemargin -.2truein
        \def\@oddfoot{\sl preliminary draft \hfil
        \rm\thepage\hfil\sl\today\quad\militarytime}
        \let\@evenfoot\@oddfoot \overfullrule 3pt
        \let\label=\draftlabel
        \let\marginnote=\draftmarginnote
   \def\@eqnnum{(\theequation)\rlap{\kern\marginparsep\tt\@eqnlabel}%
\global\let\@eqnlabel\@vacuum}  }
\def\thebibliography#1{
\vskip 0.5cm \centerline{\bf \Large References}
\list{
[\arabic{enumi}]}{\settowidth\labelwidth{[#1]}
\leftmargin\labelwidth
\advance\leftmargin\labelsep
\usecounter{enumi}}
\def\newblock{\hskip .11em plus .33em minus .07em}
\sloppy\clubpenalty4000\widowpenalty4000
\sfcode`\.=1000\relax}

\renewcommand{\theequation}{\arabic{section}.\arabic{equation}}
\renewcommand{\section}{\setcounter{equation}{0}\@startsection
{section}{1}{0mm}{-\baselineskip}{0.5\baselineskip} {\normalfont\Large\bfseries}}
\renewcommand{\subsection}{\@startsection
{subsection}{2}{0mm}{-\baselineskip}{0.5\baselineskip} {\normalfont\large\bfseries}}
\renewcommand{\subsubsection}{\@startsection
{subsubsection}{3}{0mm}{-\baselineskip}{0.5\baselineskip}
{\normalfont\normalsize\slshape}}

\begin{document}


\begin{titlepage}
\begin{flushright}
CPHT-RR087.122022, December 2022
\vspace{0.0cm}
\end{flushright}
\begin{centering}
\vspace{10mm}
{\bm\bf \Large A unimodular-like string effective description }

\vspace{10mm}

 {\bf Alex Kehagias,$^1$\footnote{kehagias@central.ntua.gr} {\bf Herv\'e Partouche$^2$}\footnote{herve.partouche@polytechnique.edu} and Nicolaos Toumbas$^3$\footnote{nick@ucy.ac.cy} 
 }

 \vspace{3mm}

$^1$  {\em Department of Physics, School of Natural Sciences, NTUA \\15780 Athens, Greece}

$^2$  {\em CPHT, CNRS, \'Ecole polytechnique, Institut Polytechnique de Paris, \\91120 Palaiseau, France}

$^3$  {\em Department of Physics, University of Cyprus, \\Nicosia 1678, Cyprus}

\end{centering}
\vspace{0.5cm}
$~$\\
\centerline{\bf\Large Abstract}\\
\vspace{-0.6cm}

\begin{quote}

We show that vanishing of the sigma-model beta-functions for the metric $g_{\mu\nu}$, antisymmetric tensor $B_{\mu\nu}$ and dilaton $\Phi$ of any string-theory model in $D$ dimensions at tree level can be obtained from a universal effective action, provided that the field combination $\sqrt{-g}e^{-2\Phi}$ is non-dynamical. The central-charge deficit term is omitted from the action and appears as an integration constant at the level of the field equations. The analysis is done at leading order in $\alpha'$, \ie at the two-derivative level in the effective description. It can be extended to incorporate higher order string-loop corrections, Ramond-Ramond forms and moduli fields. In all cases, the non-dynamical field combination turns out to be $\sqrt{-g}\, U$, where $U$ is the effective potential. When certain fluxes are fixed and the moduli fields are stabilized, no cosmological term will arise in the action. Instead, the cosmological constant will appear as an integration constant as in unimodular modifications of general relativity.     


\end{quote}

\end{titlepage}
\newpage
\setcounter{footnote}{0}
\renewcommand{\thefootnote}{\arabic{footnote}}
 \setlength{\baselineskip}{.7cm} \setlength{\parskip}{.2cm}

\setcounter{section}{0}


\section{Introduction}

Although the General Theory of Relativity has been tremendously successful in describing the gravitational interactions at the classical level, it presents various unresolved issues like for instance the cosmological constant problem. See \eg \Refs{wein, polchinski} for review. Despite considerable effort over the years, the discrepancy between theory and experiment for the value of the cosmological constant, which amounts to 120 orders of magnitude, still remains unresolved.

An interesting direction to approach the cosmological constant problem is to modify the theory of classical gravity so that the cosmological constant appears as an integration constant and {\it not as a fundamental parameter in the Lagrangian of the theory} \cite{wein,bij,zee,buch,buch2,teit,alvarez1,Alvarez:2021cxy, alvarez10,shap,garriga,blas,alvarez2,
karananas}. In particular, recall that if the determinant of the metric is treated as a non-dynamical field, the variations in the metric should be restricted to satisfy $g_{\mu\nu}\delta g^{\mu\nu}=0$ \cite{wein}. As a result, the cosmological term can be omitted from the action and the equations of motion reduce to the traceless part of the Einstein field equations,
\begin{equation}\label{Etracelss}
R_{\mu\nu}-\frac{1}{4}g_{\mu\nu}R=8\pi G \Big(T_{\mu\nu}- \frac{1}{4}g_{\mu\nu}T^{\lambda}_{\lambda}\Big).
\end{equation}
This traceless part of the equations of motion does not involve a cosmological constant term. 

Naively we seem to loose one combination of the equations of motion. However, the Bianchi identity implies that \cite{wein}
\begin{equation}
 \partial_{\mu}\Big(R+8\pi G T^{\lambda}_{\lambda}\Big)=0   ,
\end{equation}
leading to 
\begin{equation}
R+8\pi G T^{\lambda}_{\lambda}=-4 \Lambda = \rm{constant}.
\end{equation}
Combining with \Eq{Etracelss}, we obtain Einstein's equations in the usual form,
\begin{equation}
    R_{\mu\nu}-\frac{1}{2}g_{\mu\nu}R-\Lambda g_{\mu\nu}=8\pi G T_{\mu\nu},
\end{equation}
but with the cosmological constant arising as an integration constant. Notice in particular that in the absence of matter and radiation, all de Sitter backgrounds, irrespective of curvature, are solutions of \Eq{Etracelss}. Among these solutions, one must choose the one that is compatible with the observational data. Of course, still it would be desirable to have a selection mechanism among the solutions based solely on theoretical considerations and compare with the experiment.  

It is important to stress that in the approach outlined above, the formalism retains the full general coordinate covariance, except that the determinant of the metric $\sqrt{-g}$ is treated as a non-dynamical background field ($g_{\mu\nu}\delta g^{\mu\nu}=0$). We are however always free to use the gauge freedom of the theory to work in a frame where $\sqrt{-g}=1$. In such a frame, the gauge symmetry is not fixed completely. The residual gauge freedom amounts to the group of volume-preserving diffeomorphisms. 
As Weinberg emphasized~\cite{wein}, the above point of view, 
where general coordinate covariance is a symmetry, differs from other ``unimodular'' formulations of gravity, \eg \Refs{zee,buch,buch2}, in which the gauge symmetry of the theory is restricted to the group of volume-preserving diffeomorphisms. In such unimodular versions of gravity, the determinant of the metric is dynamical and behaves like a scalar field since it is invariant under volume-preserving diffeomorphisms. As a result, there is freedom to introduce arbitrary functions of this determinant in the action.  

In the framework of string theory, we use the perturbative, on-shell formulation of the theory, as well as symmetries and string dualities, to construct the low energy effective action. Indeed, at the moment, a complete off-shell formulation in terms of a string field theory action is lacking. Therefore, we cannot derive the low energy effective action directly by taking a suitable low energy or a classical limit. Rather, we construct the effective action for the massless fields in order to reproduce the string-sigma model equations, which follow from conformal invariance, and the string S-matrix. 

At the two-derivative level, however, the vanishing of the string sigma-model beta functions does not uniquely specify the form of the low energy effective action. As we will see, the alternative formulation of general relativity described above can be readily extended to the string theory context, with a certain combination of fields treated as a non-dynamical background. More specifically, we may treat a particular combination of the dilaton $\Phi$ and the determinant of the spacetime metric as a non-dynamical field and reproduce the string sigma-model equations. At tree-level, this combination turns out to be $\sqrt{-g}e^{-2\Phi}$, where $g$ is the determinant of the string-frame metric $g_{\mu\nu}$. The beta-function equations associated with $g_{\mu\nu}(x)$ and the antisymmetric tensor $B_{\mu\nu}(x)$ field follow by  extremizing the low energy effective action. The Bianchi identity can then be used to show that the beta function associated with the dilaton field $\Phi(x)$ must be a constant: $\beta^{\Phi}= \rm constant$. To recover conformal invariance, or select a proper string background, we set this integration constant to be zero.

This formalism allows us to omit non-universal terms associated with the central charge deficit from the action, in the same way the cosmological term can be omitted from the action in the case of general relativity. As a result, we may use a universal form for the low energy effective action in string frame to describe allowable string backgrounds. In the Einstein frame, the term associated with the central charge deficit appears as a dilaton-tadpole term in the action. The formalism that we develop allows us to obtain the central-charge deficit as an integration constant and omit the corresponding tadpole term from the low energy effective action. 

At the two-derivative level, the analysis can also be generalized in various ways. First, it can incorporate  quantum corrections at higher order in the string coupling $e^{\Phi}$. As we will see, it is possible to adjust the non-dynamical combination of the dilaton field and the metric determinant, order by order in the string coupling, with the low-energy effective action retaining a universal form. In fact, the action will {\it not involve any dilaton-potential term}. Nevertheless, the string equations of motion will be fully reproducible after properly incorporating the Bianchi identity and determining the field combination that must be treated as non-dynamical. The dilaton potential term, which can be derived from higher genus string amplitudes, appears instead in the field combination that is treated as non-dynamical. 

Second, extra degrees of freedom can be included in the effective description, such as Ramond-Ramond (RR) forms.  Additional scalar fields arising from the NSNS of RR sectors can also be incorporated, including those that are massless at tree level, \ie  the moduli. Denoting by $U$ the perturbative string-frame effective potential that depends on the scalar fields, the non-dynamical combination of fields in this more general case turns out to be $\sqrt{-g}\,U$. In fact, our analysis can be extended to the non-perturbative level, where string dualities and/or supersymmetry can be used to determine $U$ and the form of the low energy effective action. Since the potential term $U$ does not appear in the  effective action, even when certain fluxes are fixed and the moduli stabilized, no cosmological term will arise in the action. Rather the cosmological constant can be treated as an integration constant, as in the case of unimodular modifications of general relativity.

In~\Sect{new}, we show that at leading order in $\alpha'$, the sigma-model beta-functions associated with the metric, antisymmetric tensor and dilaton fields can be recovered from a universal action, while $\sqrt{-g}e^{-2\Phi}$ is non-dynamical and the central-charge deficit appears as an integration constant at the level of the field equations. \Sect{RRf} generalizes the analyses to include RR forms, while higher-order loop corrections in the string coupling are incorporated in~\Sect{loops}. Extra scalars, and in particular the moduli fields, are included in~\Sect{mod}. In \Sect{spec}, we show that the number of propagating degrees of freedom is not reduced even if a field combination is non-dynamical. We conclude with a discussion and present open questions in  \Sect{cl}.

\section{Another approach to the string effective action}
\label{new}
Let us briefly review the string effective action approach to worldsheet conformal invariance in string theory. We consider a string with worldsheet coordinates $(\tau,\sigma)$ moving in a $D$-dimensional spacetime. The embedding fields are denoted by $X^{\mu}(\tau,\sigma)$, $\mu\in\{0,1,\dots, D-1\}$, the spacetime metric by $g_{\mu\nu}(X)$, the antisymmetric-tensor field by $B_{\mu\nu}(X)$ and the dilaton by $\Phi(X)$. The condition for conformal invariance amounts to the vanishing of the string sigma-model $\beta$-functions \cite{gaume,frie1, Callan,frie2,sen,frad1,frad2,tsey}. The latter are then viewed as the field equations of motion associated with a spacetime effective action, which takes the form 
\be
\mathcal{S}=\frac{1}{2\kappa_0^2}\int \d^D x \sqrt{-g}
e^{-2\Phi}\left(R+4 \partial_\mu \Phi \partial^\mu \Phi -\frac{1}{12}H_{\mu\nu\rho}H^{\mu\nu\rho}+\frac{\lambda}{\alpha'}\right),
\label{act}
\ee
where $\alpha'$ is the Regge slope and $\lambda=2\delta c/3$, with $\delta c$ the central charge deficit ($\delta c=26-D$ for the example of the bosonic string).
The variation of the action (\ref{act}) yields 
\begin{align}
\delta \mathcal{S}=\frac{1}{2\kappa_0^2}\int \d^D x& \sqrt{-g}
e^{-2\Phi}\left\{
\delta g^{\mu\nu} \left(R_{\mu\nu}+2 \nabla_\mu\nabla_\nu \Phi-\frac{1}{4} H_{\mu\rho\sigma}{H_\nu}^{\rho\sigma}\right)
 +\frac{1}{2}\delta B_{\mu\nu} e^{2\Phi}\nabla_\rho \left(e^{-2\Phi}
 H^{\mu\nu\rho}\right)\right.\nonumber \\
 &\left.-\left(\frac{1}{2}g_{\mu\nu}\delta g^{\mu\nu}+2\delta \Phi\right)\left(R+4\nabla^2\Phi -4\partial_\rho \Phi \partial^\rho \Phi -\frac{1}{12} H_{\rho\sigma\tau}H^{\rho\sigma\tau} +\frac{\lambda}{\alpha'}\right)\right\}.
 \label{deltaS}
\end{align}
 Extremizing the action, $\delta \mathcal{S}=0$, leads to the vanishing of the string sigma-model $\beta$-functions 
\begin{align}
\frac{\beta^g_{\mu\nu}}{\alpha'}&=R_{\mu\nu}+2\nabla_\mu \nabla_\nu \Phi-
\frac{1}{4} H_{\mu\rho\sigma}{H_\nu}^{\rho\sigma}=0, \label{bg}\\
\frac{\beta^B_{\mu\nu}}{\alpha'}&=-\frac{1}{2}\nabla^\rho \left(e^{-2\Phi}
 H_{\mu\nu\rho}\right)=0, \label{bb}\\
 -\frac{2\beta^\Phi}{3\alpha'}&=R+4\nabla^2\Phi -4\partial_\mu \Phi \partial^\mu \Phi -\frac{1}{12} H_{\mu\nu\rho}H^{\mu\nu\rho} +\frac{\lambda}{\alpha'}=0.  \label{bf}
\end{align}

We would like to stress two points concerning the string effective action in Eq. (\ref{act}): 
\begin{itemize}

\item In \Eq{deltaS}, we see that the contributions to $\delta \S$ arising from the variations $\delta\Phi$ and $\frac{1}{2}g_{\mu\nu}\delta g^{\mu\nu}=-\delta \sqrt{-g}/\sqrt{-g}$ are proportional. Therefore, once we impose the dilaton field equation $\delta {\cal S}/\delta\Phi=0$, the variation of $\sqrt{-g}$ does not induce any contribution to $\delta\S$ or  $\beta_{\mu\nu}^g$. As a result, there is a possibility to restrict the variation of the determinant of the metric to be equal to the variation of a generic functional $A$ of all fields,  
\be
\delta \sqrt{-g}= \delta \Big[A(\Phi,g_{\mu\nu},B_{\mu\nu})\Big].  \label{gg}
\ee
Upon imposing such an extra constraint, we still obtain the same field equations (\ref{bg}),~(\ref{bb}) and~(\ref{bf}).

\item 
Another observation is that for any particular model, we have a different effective action since each model is associated with its own central-charge deficit. 
Therefore, there is no ``universal'' string effective action that describes in a unified way all string models: Each string model has his own low energy effective action, the field equations of which lead to conformal invariance when satisfied. 
\end{itemize}
The comments above seem to point towards the conclusion that  the string effective action is not unique, and the question arises if we can use a single string effective action to describe the possible models.
 
To answer this question, let us observe from \Eq{deltaS} that there is another way to achieve $\delta \mathcal{S}=0$. We can keep equations (\ref{bg}) and (\ref{bb}) but instead of Eq. (\ref{bf}), we may impose  
\be
\frac{1}{2}g_{\mu\nu}\delta g^{\mu\nu}+2\delta \Phi=0,  \label{fg}
\ee  
which is equivalent to\footnote{This condition has already been used in \cite{Perelman}.}
\be
\delta \left(\sqrt{-g}e^{-2\Phi}\right)=0.
\label{treecond}
\ee
That is, the field $\sqrt{-g}e^{-2\Phi}$ is treated to be non-dynamical with vanishing variation. 
Notice that even though $\sqrt{-g}e^{-2\Phi}$ is treated as a background field, the theory retains the full diffeomorphism invariance. This is similar to Weinberg's approach outlined in the introduction for the cosmological constant problem, where the determinant of the Einstein-frame metric is treated to be non-dynamical~\cite{wein}. In the case at hand, one can use the gauge freedom of the theory to set $\sqrt{-g}e^{-2\Phi}=1$. 
This is simply a gauge choice. This condition does not fix the gauge freedom completely. There is a residual gauge symmetry comprising volume-preserving (transverse) diffeomorphisms. Indeed, the condition $\sqrt{-g}e^{-2\Phi}=1$ is invariant under diffeomorphisms $x^\mu\to x^\mu+\xi^\mu$ that satisfy $\partial_\mu \xi^\mu=1$. 

To recap, in the above approach, the equations of motion to be satisfied are those associated with the metric $g_{\mu\nu}$ and the antisymmetric tensor $B_{\mu\nu}$,  
\begin{align}
&R_{\mu\nu}+2\nabla_\mu \nabla_\nu \Phi-
\frac{1}{4} H_{\mu\nu\rho}H^{\mu\nu\rho}=0, \label{bg1}\\
&\nabla^\rho \left(e^{-2\Phi}
 H_{\mu\nu\rho}\right)=0,  \label{bb1}
 \end{align}
together with condition (\ref{treecond}). Observe that the central charge deficit does not appear in \Eqs{bg1}, (\ref{bb1}) and~(\ref{treecond}). As a result, we are free to start with the universal action
  \be
\widehat{\mathcal{S}}=\frac{1}{2\kappa_0^2}\int \d^D x \sqrt{-g}
e^{-2\Phi}\left(R+4 \partial_\mu \Phi \partial^\mu \Phi -\frac{1}{12}H_{\mu\nu\rho}H^{\mu\nu\rho}\right),
\label{act1}
\ee
subject to (\ref{treecond}). A question then arises: Do the extrema of $\widehat \S$ describe conformal string backgrounds? For this to be true, we need $\beta^\Phi=0$. However, it is known that the vanishing of $\beta^g_{\mu\nu}$ and $\beta^B_{\mu\nu}$ leads to \cite{Callan}
\be
\beta_{\mu\nu}^g=\beta_{\mu\nu}^B=0~~~\Longrightarrow~~~~\nabla^\mu\beta^\Phi=0.
\ee
To see this, recall that by using the Bianchi identity, we obtain  
\be
 0=\nabla^\mu \beta^g_{\mu\nu}=-2\nabla_\nu \left(-\frac{1}{2}\nabla^2 \Phi+ \partial_\mu \Phi \partial^\mu \Phi-\frac{1}{24} H_{\mu\nu\rho} H^{\mu\nu\rho}\right) 
\ee
and so
\be
-\,\frac{1}{2}\nabla^2 \Phi+ \partial_\mu \Phi \partial^\mu \Phi-\frac{1}{24} H_{\mu\nu\rho} H^{\mu\nu\rho}=c_0,
\ee
where $c_0$ is a constant. Choosing appropriately $c_0=\delta c/6\alpha'$, this is equivalent to Eq. (\ref{bf}) \ie 
\be
\beta^\Phi=0.
\ee

To conclude this subsection, let us stress again that the central-charge deficit does not appear in the string effective action (\ref{act1}). Instead, it arises as an integration constant, as in the case of the cosmological constant in unimodular gravity. 

\subsection{Generalization}

To check whether the string effective action (\ref{act1}) is special, 
let us consider the most general diff invariant action for the metric and dilaton (ignoring the antisymmetric tensor for simplicity), where we keep terms with two derivatives only and no dilaton potential. Without loss of generality, it can be written as 
 \be
 \widehat\S=\frac{1}{2\kappa_0^2}\int \d^D x \sqrt{-g}F(\Phi)\bigg(R+4 \partial_\mu \Phi\partial^\mu \Phi
 \bigg),   \label{eaa}
 \ee
where $F(\Phi)$ is an arbitrary function of $\Phi$. Indeed, 
even if we start with an action involving a generic sigma-model kinetic term $G(\Phi)\partial_\mu\Phi\partial^\mu\Phi$, where $G$ and $F$ are different functions, we can make a field redefinition of~$\Phi$ to bring the action to the form \ref{eaa}.\footnote{This is true for a one-dimensional sigma-model.} We would like to determine under which condition the contribution to $\delta\widehat S$ from the variation of $\sqrt{-g}$ vanishes, once the equation of motion for $\Phi$ is satisfied. Indeed, when this occurs, the constraint that $\delta\sqrt{-g}$ should be proportional to $\delta\Phi$, as in equation (\ref{fg}), is harmless, since the 
equation of motion for $g_{\mu\nu}$ will not be affected. 

The variation of the action is  
\begin{align}
 \delta\widehat\S=\frac{1}{2\kappa_0^2}\int \d^D x  \sqrt{-g}F(\Phi)
& \bigg\{ \delta g^{\mu\nu }\left(R_{\mu\nu}+4 \nabla_\mu \Phi\nabla_\nu \Phi-\frac{\nabla_\mu \nabla_\nu F}{F}\right) \nonumber \\
 &-\frac{1}{2}g_{\mu\nu}\delta g^{\mu\nu}\bigg(R+4 \partial_\rho \Phi\partial^\rho \Phi-2\,\frac{\nabla^2 F}{F}\bigg)  \nonumber\\
 &+\frac{F'}{F}\delta \Phi
 \bigg(R+4\partial_\mu \Phi \partial^\mu \Phi-8\, \frac{\partial_\mu\big(F\partial^\mu \Phi\big)}{F'}\bigg) \bigg\}, \label{ssss}
\end{align}
where $F'=d F/d\Phi$. From the last two lines of \Eq{ssss}, we see that in order for the contribution proportional to $g_{\mu\nu} \delta g^{\mu\nu}$ to vanish when the contribution proportional to $\delta \Phi$ is zero, we must have 
\be
-2\,\frac{\nabla^2 F}{F}=-8 \,\frac{\partial_\mu\big(F\partial^\mu \Phi\big)}{F'}.
\ee
This is satisfied when 
\begin{eqnarray}
\frac{F'}{F}=4\, \frac{F}{F'} ~~~~~\mbox{and}~~~~~\frac{F''}{F}=4,
\end{eqnarray}
leading to 
\be
F(\Phi)=e^{\pm 2 \Phi}. 
\ee
Choosing the minus sign, we get (\ref{act1}) (for the metric and dilaton). Of course, the choice of plus sign is also possible, but it is equivalent under the trivial field redefinition $\Phi\to -\Phi$. 
In other words, the string effective action~(\ref{act1}) is the unique action of the form (\ref{eaa}), 
for which the contribution to $\delta\widehat S$ from the variation of the determinant of the metric vanishes when the dilaton equation holds.

\subsection{An application} 
\label{WBH}

In the case of vanishing antisymmetric tensor field ($B_{\mu\nu}=0$), we have seen that conformal invariance is guaranteed when 
\be
R_{\mu\nu}+2\nabla_\mu\nabla_\nu \Phi=0, \label{eq1}
\ee
subject to  the condition of Eq.(\ref{treecond}). As we have already remarked, we may fix the gauge 
\be
\sqrt{-g} e^{-2\Phi}=e^{-2\Phi_0},\label{gf}
\ee
where $\Phi_0$ is an arbitrary constant. In that case,  Eq. (\ref{eq1}) takes the form
\be
R_{\mu\nu}+\nabla_\mu\bigg(\frac{1}{\sqrt{-g}}\nabla_\nu\sqrt{-g}\bigg)=0. \label{eq2}
\ee
Using the definition of the Ricci tensor in terms of  Christoffel symbols, it is then straightforward to verify that the above equation can be written as 
\be
K_{\mu\nu}=0, \qquad \where \qquad K_{\mu\nu}=\partial_\kappa \Gamma^\kappa_{\mu\nu}-\Gamma^\kappa_{\nu\lambda}\Gamma^\lambda_{\kappa\mu}.
\label{eq3}
\ee
One may worry about the fact that  \Eq{eq3} is not covariant. Indeed, under a general coordinate transformation $x^\mu\to \tilde x^\mu(x^\nu)$, $K_{\mu\nu}$ transforms as 
\be
K_{\mu\nu}\longrightarrow K_{\mu'\nu'}=\frac{\partial x^\mu}{\partial x^{\mu'}}
\frac{\partial x^\nu}{\partial x^{\nu'}}   K_{\mu\nu}+
\frac{\partial x^\mu}{\partial x^{\mu'}} 
\frac{\partial x^\nu}{\partial x^{\nu'}}\nabla_\mu\partial_\nu
\ln\left(\det\frac{\partial x^\alpha}{\partial x^{\alpha'}}\right).
%
%
\ee
Therefore, although $K_{\mu\nu}$ is not a tensor under the full diffeomorphism group, it is a tensor under volume-preserving diffeomorphisms ($\det\partial x^\alpha/\partial x^{\alpha'}=1)$, as expected, since the gauge we have chosen has the residual volume-preserving diffeomorphism invariance. 

Let us now study an example. We will consider a two-dimensional Euclidean space, with metric~
\be
\d s^2=f(r) \Big( \d r^2+r^2 \d \theta^2\Big), \label{met1}
\ee
where $f(r)$ is to be determined. In that case, Eq. (\ref{eq2}) reduces to
\be
\frac{1}{2} f''-\frac{f'{}^2}{f}-\frac{f'}{r}-\frac{f}{r^2}=0,
\ee
whose solution is   
\be
f(r)=\frac{f_0}{r^2}\,\frac{1}{1-\dfrac{M}{r}}, \label{sol1}
\ee
where $f_0$ and $M$ are constants. Using \Eq{gf}, the dilaton is given by 
\be
\Phi=\frac{1}{2}\ln \sqrt{g}+\Phi_0=\frac{1}{2}\ln\Big(rf(r)\Big) +\Phi_0. \label{dil1}
\ee
The constant $f_0$ determines the associated two-dimensional CFT. In particular, 
the metric (\ref{met1}), (\ref{sol1}) and the dilaton (\ref{dil1}) describe the Witten black hole if $k=8f_0$ is identified with the level $k$ of the $SU(2)$ current algebra, since it can be readily verified that $\Phi$ satisfies  
\be
-\frac{1}{2}\nabla^2 \Phi +\nabla_\mu \Phi \nabla^\mu \Phi=\frac{1}{4f_0}.
\ee

Interestingly, in our approach, the level $k$ appears in the equations as an integration constant. Since it is not present in the action~(\ref{act1} subject to the constraint~(\ref{treecond}), a universal effective description is used to encompass the string solutions irrespectively of $k$.


\section{Inclusion of  RR fields}
\label{RRf}

 In type II string theory, one should include RR-fields in the effective action (\ref{act}). We will omit the NS-NS two-form and consider only the NS-NS metric and  dilaton, along with a RR one-form for simplicity:
 \be
     \mathcal{S}=\int \mathrm{d}^{D} x \sqrt{-g}\left\{\frac{1}{2 \kappa_{0}^{2}} e^{-2 \Phi}\left(R+4 \partial_{\mu} \Phi \partial^{\mu}\Phi+\frac{\lambda}{\alpha'}\right)-\frac{1}{4} F_{\mu \nu} F^{\mu \nu}\right\}.
     \label{RR-ac}
 \ee
The variation of this action  turns out to be 
\begin{align}
\delta \mathcal{S}=\int \mathrm{d}^{D} x \sqrt{-g}\left\{\frac{1}{2 \kappa_{0}^{2}} e^{-2 \Phi}\right. & \bigg[\delta g^{\mu \nu}\Big(R_{\mu \nu}+2 \nabla_{\mu} \nabla_{\nu} \Phi\Big)\nonumber  \\
&-\left(\frac{1}{2} g_{\mu \nu} \delta g^{\mu \nu}+2 \delta \Phi\right)\left(R+4 \nabla^{2} \Phi-4 \partial_{\rho} \Phi \partial^{\rho} \Phi +{\lambda\over\alpha'}\right)\bigg] \nonumber  \\
& \left.-\;\frac{1}{2} \delta g^{\mu \nu}\left(F_{\mu \rho} F_{\nu}^{\rho}-\frac{1}{4} g_{\mu \nu} F_{\kappa \lambda} F^{\kappa \lambda}\right)+\delta A_{\nu} \nabla_{\mu} F^{\mu \nu}\right\},
\end{align}
leading to the following equations of motion:
\begin{align}
& R_{\mu \nu}+2 \nabla_{\mu} \nabla_{\nu} \Phi=\kappa_{0}^{2} e^{2 \Phi}\left(F_{\mu \rho} F_{\nu}^{\rho}-\frac{1}{4} g_{\mu \nu} F_{\kappa \lambda} F^{\kappa \lambda}\right) \label{eqr1}\\
& \nabla_{\mu} F^{\mu \nu}=0,\label{eqr2} \\
& R+4 \nabla^{2} \Phi-4 \partial_{\mu} \Phi \partial^{\mu} \Phi+\frac{\lambda}{\alpha'}=0 . \label{eqr3}
\end{align}

Let us now consider the alternative action
 \be
 \widehat\S=\int \mathrm{d}^{D} x \sqrt{-g}\left\{\frac{1}{2 \kappa_{0}^{2}} e^{-2 \Phi}\Big(R+4 \partial_{\mu} \Phi \partial^{\mu}\Phi\Big)-\frac{1}{4} F_{\mu \nu} F^{\mu \nu}\right\},
     \label{RR-ac2}
 \ee
 where there is no reference to $\lambda$. Imposing the constraint of \Eq{fg}, 
the only field equations we have to consider are \Eqs{eqr1} and~(\ref{eqr2}), 
\begin{align}
& R_{\mu \nu}+2 \nabla_{\mu} \nabla_{\nu} \Phi=\kappa_{0}^{2} e^{2 \Phi}\left(F_{\mu \rho} F_{\nu}^{\rho}-\frac{1}{4} g_{\mu \nu} F_{\kappa \lambda} F^{\kappa \lambda}\right), \label{eqr11}\\
& \nabla_{\mu} F^{\mu \nu}=0.\label{eqr22}
\end{align}
Hence, it seems again that Eq. (\ref{eqr3}) for the dilaton is missing. However, we will show below that the dilaton equation of motion follows from  Eqs. (\ref{eqr11}) and (\ref{eqr22}), and that the central-charge deficit, which is proportional to $\lambda$, appears as an integration constant. 

To see this, let us note that from 
Eq. (\ref{eqr11}) we find that
\be
    \nabla^{\mu}\Big(R_{\mu \nu}+2 \nabla_{\mu} \nabla_{\nu} \Phi\Big)=\kappa_{0}^{2} T_{\mu \nu} \nabla^{\mu} e^{2 \Phi}+\kappa_{0}^{2} e^{2 \Phi} \nabla^{\mu} T_{\mu \nu}, \label{eqr4}
\ee
where 
\be
    T_{\mu \nu}=F_{\mu \rho} F_{\nu}^{\rho}-\frac{1}{4} g_{\mu \nu} F_{\kappa \lambda} F^{\kappa \lambda}.
\ee
Using  Eq. (\ref{eqr22}) (or, equivalently, invariance under $\delta g_{\mu\nu}=\nabla_\mu \xi_\nu+\nabla_\nu \xi_\mu$), we get that $T_{\mu\nu}$ is conserved: $\nabla^\mu T_{\mu\nu}=0$.  Then, by using the identity 
\be
    \nabla^{\mu} \nabla_{\nu} \nabla_{\mu} \Phi=\nabla_{\nu} \nabla^{2} \Phi+R_{\mu \nu} \nabla^{\mu} \Phi,
\ee
we find that Eq. (\ref{eqr4}) can be written as follows
\be
    \nabla_{\nu}\left(\frac{1}{2} R+2 \nabla^{2} \Phi-2 \partial_{\mu} \Phi \partial^{\mu} \Phi\right)=0.
\ee
Therefore, we obtain the desired equation 
\be
R+4 \nabla^{2} \Phi-4 \nabla_{\mu} \Phi \nabla^{\mu} \Phi   + \frac{\lambda}{\alpha'}=0,
\ee
where $\lambda$ now appears as an integration constant. 
In other words, the action~(\ref{RR-ac2}), where the field combination $\sqrt{-g}e^{-2\Phi}$
is non-dynamical (${1\over 2}g_{\mu \nu} \delta g^{\mu \nu}+2 \delta \Phi=0$), yields the same equations of motion as the original string effective action (\ref{RR-ac}). 

\section{Extension to higher string-loops}
\label{loops}

After showing that $\sqrt{-g} e^{-2\Phi}$ can be non-dynamical including RR fields (which are string one-loop effects compared to the tree-level string action), one may wonder if the analysis can be extended to higher-order string loops. In other words, we would like to investigate if it is possible to keep a field combination of $\sqrt{-g}$ and $\Phi$ non-dynamical when higher string loops are included, or if this feature is a particularity of the tree-level string effective action. One may expect that the combination $\sqrt{-g} e^{-2\Phi}$ may need to be corrected by higher-order contributions in string coupling. To see what is going on, let us incorporate string-loop corrections in the string effective action at the two-derivative level. The action takes the general form 
\be
     \mathcal{S}=\frac{1}{2 \kappa_{0}^{2}} \int \mathrm{d}^{D} x \sqrt{-g} \bigg(F(\Phi)R+G(\Phi) \partial_{\mu} \Phi \partial^{\mu} \Phi-U(\Phi)\bigg),
    \label{ac-l}
\ee
where $F(\Phi)$ and $G(\Phi)$ admit a string loop expansion of the form
\begin{align}
    F(\Phi)&=e^{-2\Phi}+c_0 +c_1 e^{2\Phi}+c_2 e^{4\Phi}+\cdots \nonumber \\
     G(\Phi)&=4e^{-2\Phi}+d_0 +d_1 e^{2\Phi}+d_2 e^{4\Phi}+\cdots\, .
\end{align}
We have also included a dilaton potential, which can be expanded in a similar way
\be
    U(\Phi)= -\frac{\lambda}{\alpha'} e^{-2\Phi}+u_0 +u_1 e^{2\Phi}+u_2 e^{4\Phi}+\cdots \, . 
\ee

It is convenient at this point to transform the above action to the Einstein frame \via the following conformal transformation,
\be
    g_{\mu\nu}=F^{-\frac{2}{D-2}}\bar{g}_{\mu\nu},  \label{conf}
\ee
which leads to
\be
     \mathcal{S}=\frac{1}{2 \kappa_{0}^{2}} \int \mathrm{d}^{D} x \sqrt{-\bar g} \bigg(\bar R+H(\Phi) \partial_{\mu} \Phi \partial^{\mu} \Phi-\bar U(\Phi)\bigg).
    \label{ac-ei}
\ee
In this formula, we have defined  
\be
    H(\Phi)=-\frac{D-1}{D-2}\left(\frac{F'}{F}\right)^2+\frac{G}{F},\qquad \bar U(\Phi)=
    F^{-\frac{D}{D-2}}\, U
    \label{potE}
\ee
and a prime denotes derivative with respect to the dilaton $\Phi$. The variation of the action  (\ref{ac-ei}) turns out to be 
\begin{align}
    \delta \mathcal{S}&= \frac{1}{2 \kappa_{0}^{2}}\int \mathrm{d}^{D} x \sqrt{ -\bar g}\bigg[  \delta \bar g^{\mu \nu}\left(\bar R_{\mu \nu}+H\partial_\mu\Phi\partial_\nu\Phi\right)-\frac{1}{2} \bar g_{\mu \nu} \delta \bar g^{\mu \nu}\left(\bar R+H(\partial\Phi)^2-\bar U\right)\nonumber \\
    &\hspace{3cm}
 +    \delta \Phi\left(H' (\partial \Phi)^2-2\bar\nabla_\mu(H\partial^\mu \Phi)-\bar U'\right)\bigg],
\end{align}
where $\bar \nabla_\mu$ stands for the Levi-Civita connection associated with the Einstein-frame metric. 
Therefore, the equations of motion are 
\begin{align}
   &  \bar R_{\mu\nu}+H\partial_\mu\Phi\partial_\nu\Phi-\frac{1}{2}  \bar g_{\mu \nu}\left(\bar R+H(\partial\Phi)^2-\bar U\right)=0,
   \label{eq01}\\
   &H' (\partial \Phi)^2-2\bar\nabla_\mu(H\partial^\mu \Phi)-\bar U'=0.  \label{eq02}
\end{align}

We look for a string effective action {\it without any dilaton potential}, which takes the following form in the Einstein frame 
\be
     \widehat\S=\frac{1}{2 \kappa_{0}^{2}} \int \mathrm{d}^{D} x \sqrt{-\bar g} \bigg(\bar R+H(\Phi) \partial_{\mu} \Phi \partial^{\mu} \Phi\bigg),
    \label{ac-eibis}
\ee
subject to the condition
\be
   K(\Phi)^{-1} \delta\Phi=  \bar g_{\mu\nu}\delta \bar g^{\mu\nu}, \label{cc1}
\ee
for some function $K(\Phi)$. The latter should admit a perturbative expansion of the form
\be
    K(\Phi)=- {1\over 4}+k_1 e^{2\Phi}+k_2e^{4\Phi}+\cdots\,.
\ee
Given condition (\ref{cc1}), we find that the variation of the action is now given by
\begin{align}
    \delta \widehat\S= \frac{1}{2 \kappa_{0}^{2}} \int \mathrm{d}^{D} x \sqrt{-\bar g}&\bigg[ \delta \bar g^{\mu \nu}\left(\bar R_{\mu \nu}+H\partial_\mu\Phi\partial_\nu\Phi\right)\nonumber \\
    &   -\frac{1}{2} \bar g_{\mu \nu} \delta \bar g^{\mu \nu}\left(\bar R+H(\partial\Phi)^2+2K\Big(2 \bar\nabla_\rho(H \partial^\rho \Phi)-H' (\partial\Phi)^2\Big)\right)\bigg],
\end{align}
leading to a single equation of motion
\be
    R_{\mu\nu}+H\partial_\mu\Phi\partial_\nu\Phi-\frac{1}{2}   \bar g_{\mu \nu}\left(\bar R+H(\partial\Phi)^2+2K\Big(2 \bar\nabla_\rho(H \partial^\rho \Phi)-H' (\partial\Phi)^2\Big)\right)=0.
   \label{eq011}
\ee

We would now like, by appropriately choosing $K(\Phi)$ and utilizing the Bianchi identity, to recover the dilaton equation (\ref{eq02}). Acting with $\nabla^\mu$ on both sides of Eq. (\ref{eq011}), we obtain 
\be
 \bar\nabla^\mu \Big(H\partial_\mu\Phi\partial_\nu\Phi\Big)-\frac{1}{2}  \bar\nabla_\nu \bigg\{H(\partial\Phi)^2+2K\Big(2 \bar\nabla_\mu(H \partial^\mu \Phi)-H' (\partial\Phi)^2\Big)\bigg\}=0,
\ee
leading to 
\be
   \bigg\{\bar\nabla^\mu \Big(H\partial_\mu\Phi\Big)-\frac{1}{2}H'(\partial \Phi)^2\bigg\}\partial_\nu \Phi- \bar\nabla_\nu \bigg\{2KV\, V^{-1}\Big( \bar\nabla_\mu(H \partial^\mu \Phi)-\frac{1}{2}H' (\partial\Phi)^2\Big)\bigg\}=0, \label{eqss}
\ee
where we have introduced a function $V(\Phi)$ to be chosen suitably. Indeed,  
the above equation can be written as 
\begin{align}
    \bigg\{\bar\nabla^\mu \Big(H\partial_\mu\Phi\Big)-\frac{1}{2}H'(\partial \Phi)^2\bigg\}\bigg(\partial_\nu \Phi-&2V^{-1}\partial_\nu ( KV)\bigg)\nonumber \\
   &-2 K V  \bar\nabla_\nu \bigg\{ V^{-1}\Big( \bar\nabla_\mu(H \partial^\mu \Phi)-\frac{1}{2}H' (\partial\Phi)^2\Big)\bigg\}=0
\end{align}
and for 
\be
\partial_\nu \Phi-2V^{-1}\partial_\nu ( KV)=0, \label{kdv}
\ee
it simplifies to 
\be
   \bar\nabla_\nu \bigg\{V^{-1}\Big( \bar\nabla_\mu(H \partial^\mu \Phi)-\frac{1}{2}H' (\partial\Phi)^2\Big)\bigg\}=0,
\ee
or 
\be
\bar\nabla_\mu(H \partial^\mu \Phi)-\frac{1}{2}H' (\partial\Phi)^2=\nu V,
\label{V0}
\ee
where $\nu$ is an integration constant. Comparing Eqs. (\ref{V0}) and 
(\ref{eq02}), we see that they coincide if
\begin{equation}
    V=-\frac{1}{2\nu} \bar U'. 
\end{equation}
Then Eq. (\ref{kdv}) can be  written as 
\be
\partial_\nu \left(\bar U-2 K\bar U'\right)=0, \label{kdv1}
\ee
which specifies $K$ to be 
\begin{equation}
   K=\frac{\bar U+\kappa}{2\, \bar U'}, \label{K}
\end{equation}
where $\kappa$ is an integration constant. However, using this result and the expression of $V$, \Eq{V0} becomes
\be
2K\Big[H' (\partial\Phi)^2-2\bar\nabla_\mu(H \partial^\mu \Phi)\Big] \!= \bar U+\kappa .
\ee
As a result, for \Eq{eq011} to also be identical to the metric equation of motion (\ref{eq01}), we must have $\kappa=0$. This allows to write \Eq{cc1} as 
\begin{equation}
{ \bar U'\over \bar U}\,   \delta\Phi=  {1\over 2}\, \bar g_{\mu\nu}\delta \bar g^{\mu\nu}, 
\end{equation}
which leads to 
\be
\delta \Big[\bar U(\Phi)\, \sqrt{-\bar g}\Big]\!=0.
\ee
Using the relations between the string-frame  and Einstein-frame  metrics and potentials, \Eqs{conf} and~(\ref{potE}), this is equivalent to having
\be
\delta \Big[U(\Phi)\, \sqrt{-g}\Big]\!=0,
\ee
which generalizes the tree-level condition in \Eq{treecond}.



\section{Inclusion of scalar fields}
\label{mod}

At tree level, string theory contains moduli fields that need to be included in the effective description at low energies. In general, the moduli acquire a potential at higher-loop levels, which may lead to their stabilization. However, the alternative action (\ref{ac-eibis}) we have considered for the metric and dilaton field does not contain any scalar potential. So it is not clear how 
the extra moduli 
may be stabilized. In the following, we show that our effective description captures correctly the dynamics of moduli fields and, more generally, the dynamics of scalar fields with tree-level potential. 

To begin with, let us consider the standard string effective action for the metric and $N+1$ scalars fields $\Phi\equiv(\Phi^0, \dots,\Phi^N)$, where $\Phi^0$ stands for the dilaton. We assume non-trivial sigma-model kinetic terms and a potential,
 \be
     \mathcal{S}=\frac{1}{2 \kappa_{0}^{2}} \int \mathrm{d}^{D} x \sqrt{-g} \bigg(F(\Phi)R+G_{ij}(\Phi) \partial_{\mu} \Phi^i \partial^{\mu} \Phi^j-U(\Phi)\bigg),
    \label{ac-l2}
\ee
where $i\in\{0,\dots,N\}$ and 
\begin{align}
    F(\Phi)&=e^{-2\Phi^0}+c_0(\Phi^1,\dots,\Phi^N) +c_1(\Phi^1,\dots,\Phi^N) e^{2\Phi^0}+\cdots \nonumber \\
     G_{ij}(\Phi)&=d_{ij,-1}(\Phi^1,\dots,\Phi^N)e^{-2\Phi^0}+d_{ij,0}(\Phi^1,\dots,\Phi^N) +d_{ij,1}(\Phi^1,\dots,\Phi^N) e^{2\Phi^0}+\cdots \nonumber \\
    U(\Phi)&= u_{-1}(\Phi^1,\dots,\Phi^N)e^{-2\Phi^0}+u_0(\Phi^1,\dots,\Phi^N) +u_1(\Phi^1,\dots,\Phi^N) e^{2\Phi}+\cdots ~. 
\end{align}
In terms of the Einstein metric $\bar g_{\mu\nu}$, given by 
\be
    g_{\mu\nu}=F^{-\frac{2}{D-2}}\bar{g}_{\mu\nu},  \label{conf22}
\ee
the action becomes 
\be
     \mathcal{S}=\frac{1}{2 \kappa_{0}^{2}} \int \mathrm{d}^{D} x \sqrt{-\bar g} \bigg(\bar R+H_{ij}(\Phi) \partial_{\mu} \Phi \partial^{\mu} \Phi-\bar U(\Phi)\bigg),
    \label{ac-ei2}
\ee
where 
\be
    H_{ij}(\Phi)=-\frac{D-1}{D-2}\,\frac{\partial_iF}{F}\,\frac{\partial_jF}{F}+\frac{G_{ij}}{F},\qquad \bar U(\Phi)=
    F^{-\frac{D}{D-2}}\, U,
    \label{potE2}
\ee
and $\partial_i$ denotes derivative with respect to $\Phi^i$. Varying the action, we obtain 
\begin{align}
    \delta \mathcal{S}= \frac{1}{2 \kappa_{0}^{2}}\int \mathrm{d}^{D} x \sqrt{ -\bar g}&\bigg[  \delta \bar g^{\mu \nu}\left(\bar R_{\mu \nu}+H_{ij}\partial_\mu\Phi^i\partial_\nu\Phi^j\right)-\frac{1}{2} \bar g_{\mu \nu} \delta \bar g^{\mu \nu}\left(\bar R+H_{ij}\partial_\rho\Phi^i\partial^\rho\Phi^j-\bar U\right)\nonumber \\
    & +    \delta \Phi^k\left(\partial_kH_{ij} \partial_\mu \Phi^i\partial^\mu\Phi^j-2\bar\nabla_\mu(H_{ki}\partial^\mu \Phi^i)-\partial_k\bar U\right)\bigg],
\end{align}
from which we find the equations of motion
\begin{align}
   &  \bar R_{\mu\nu}+H_{ij}\partial_\mu\Phi^i\partial_\nu\Phi^j-\frac{1}{2}   \bar g_{\mu \nu}\left(\bar R+H_{ij}\partial_\rho\Phi^i\partial^\rho\Phi^j-\bar U\right)=0,
   \label{eq012}\\
   &\partial_kH_{ij} \partial_\mu\Phi^i\partial^\mu \Phi^j-2\bar\nabla_\mu(H_{ki}\partial^\mu \Phi^i)-\partial_k\bar U=0.  \label{eq022}
\end{align}

Let us now look for an alternative effective description, where the string-frame action is identical to $\S$ in \Eq{ac-l2} {\it but without any scalar potential}. In the Einstein frame, the action takes the form 
\be
     \widehat \S=\frac{1}{2 \kappa_{0}^{2}} \int \mathrm{d}^{D} x \sqrt{-\bar g} \bigg(\bar R+H_{ij}(\Phi) \partial_{\mu} \Phi^i \partial^{\mu} \Phi^j\bigg).
    \label{ac-ei22}
\ee
As before, the latter has to be subjected to some condition on the fields, which at first order in fluctuations takes a linear form 
\be
L_i  (\Phi) \delta\Phi^i=  \bar g_{\mu\nu}\delta \bar g^{\mu\nu}. \label{cc12}
\ee
This is the most general ansatz compatible with 
diff invariance. The functions $L_i(\Phi)$ to be determined are expected to admit expansions of the form
\be
    L_i(\Phi)=\ell_{i,0}(\Phi^1,\dots,\Phi^N)+\ell_{i,1}(\Phi^1,\dots,\Phi^N) e^{2\Phi}+\cdots\, .
\ee
Using the constraint~(\ref{cc12}), the variation of $\S$ can be written in terms of the variations of the metric and of only $N$ scalar fields among the $N+1$ available. For instance, eliminating the variation of $\Phi^{i_0}$ for some $i_0\in\{0,\dots,N\}$, one way of writing the variation of the action is  
\begin{align}
    \delta \widehat \S= &\,\frac{1}{2 \kappa_{0}^{2}} \int \mathrm{d}^{D} x \sqrt{-\bar g}\bigg[ \delta \bar g^{\mu \nu}\left(\bar R_{\mu \nu}+H_{ij}\partial_\mu\Phi^i\partial_\nu\Phi^j\right)\nonumber \\
    &
   -\frac{1}{2} \bar g_{\mu \nu} \delta \bar g^{\mu \nu}\left(\bar R+H_{ij}\partial_\rho\Phi^i\partial^\rho\Phi^j+2L^{-1}_{i_0}\Big(2 \bar\nabla_\rho(H_{i_0i} \partial^\rho \Phi^i)-\partial_{i_0}H_{ij} \partial_\rho\Phi^i\partial^\rho\Phi^j\Big)\right)\nonumber \\
   &+ \delta\Phi^\hi\Big[\partial_\hi H_{ij}\partial_\mu\Phi^i\partial^\mu\Phi^j-2\bar\nabla_{\mu}(H_{\hi i}\partial^\mu\Phi^i)-{L_\hi\over L_{i_0}}\Big(\partial_{i_0} H_{ij}\partial_\mu\Phi^i\partial^\mu\Phi^j-2\bar\nabla_\mu(H_{i_0 i}\partial^\mu\Phi^i)\Big)\Big]\bigg],
\end{align}
where $\hi\in\{0,\dots,N\}\backslash\{i_0\}$.
In that case, the equations of motion for the metric and scalar fields $\Phi^\hi$ read
\begin{align}
    &R_{\mu\nu}+H_{ij}\partial_\mu\Phi^i\partial_\nu\Phi^j-\frac{1}{2}   \bar g_{\mu \nu}\bigg(\bar R+H_{ij}\partial_\rho\Phi^i\partial^\rho\Phi^j\nonumber \\
    &\hspace{5.2cm}+2L^{-1}_{i_0}\Big(2 \bar\nabla_\rho(H_{i_0i} \partial^\rho \Phi^i)-\partial_{i_0}H_{ij} \partial_\rho\Phi^i\partial^\rho\Phi^j\Big)\bigg)=0,
   \label{eq0112}\\
   &\partial_\hi H_{ij}\partial_\mu\Phi^i\partial^\mu\Phi^j-2\bar\nabla_{\mu}(H_{\hi i}\partial^\mu\Phi^i)-{L_\hi\over L_{i_0}}\Big(\partial_{i_0} H_{ij}\partial_\mu\Phi^i\partial^\mu\Phi^j-2\bar\nabla_\mu(H_{i_0 i}\partial^\mu\Phi^i)\Big)=0.\label{ss}
\end{align}

Let us first derive from \Eqs{eq0112} and~(\ref{ss}) the equation for $\Phi^{i_0}$, which is \Eq{eq022} for $k=i_0$.  Applying the covariant derivative $\bar \nabla^\mu$ on \Eq{eq0112}, we obtain
\begin{align}
\bar\nabla^\mu \Big(H_{ij}&\partial_\mu\Phi^i\partial_\nu\Phi^j\Big)\nonumber \\
&-\frac{1}{2}  \bar\nabla_\nu \bigg\{H_{ij}\partial_\mu\Phi^i\partial^\mu\Phi^j+2L^{-1}_{i_0}\Big(2 \bar\nabla_\mu(H_{i_0i} \partial^\mu \Phi^i)-\partial_{i_0}H_{ij} \partial_\mu\Phi^i\partial^\mu\Phi^j\Big)\bigg\}=0,
\end{align}
which can be rewritten as
\begin{align}
   \bigg\{\bar\nabla^\mu \Big(H_{ki}\partial_\mu\Phi^i\Big)&-\frac{1}{2}\partial_kH_{ij}\partial_\mu \Phi^i\partial^\mu\Phi^j\bigg\}\partial_\nu \Phi^k\nonumber \\
   &- \bar\nabla_\nu \bigg\{2L^{-1}_{i_0}V_{i_0} \, V^{-1}_{i_0}\Big( \bar\nabla_\mu(H_{i_0i} \partial^\mu \Phi^i)-\frac{1}{2}\partial_{i_0}H_{ij} \partial_\mu\Phi^i\partial^\mu\Phi^j\Big)\bigg\}=0, \label{eqss2}
\end{align}
where the function $V_{i_0}(\Phi)$ can be chosen arbitrarily. This yields
\begin{align}
   \bigg\{\bar\nabla^\mu \Big(H_{i_0i}\partial_\mu\Phi^i\Big)&-\frac{1}{2}\partial_{i_0}H_{ij}\partial_\mu \Phi^i\partial^\mu\Phi^j\bigg\}\bigg(\partial_\nu \Phi^{i_0}-2V_{i_0}^{-1}\partial_\nu (L_{i_0}^{-1}V_{i_0})\bigg)\nonumber \\
   &+\bigg\{\bar\nabla^\mu \Big(H_{\hi i}\partial_\mu\Phi^i\Big)-\frac{1}{2}\partial_{\hi}H_{ij}\partial_\mu \Phi^i\partial^\mu\Phi^j\bigg\}\partial_\nu\Phi^\hi\nonumber \\
   &-2 L_{i_0}^{-1} V_{i_0}  \bar\nabla_\nu \bigg\{ V_{i_0}^{-1}\Big( \bar\nabla_\mu(H_{i_0i} \partial^\mu \Phi^i)-\frac{1}{2}\partial_{i_0} H_{ij} \partial_\mu\Phi^i\partial^\mu\Phi^j\Big)\bigg\}=0.
\end{align}
Using \Eq{ss}, the second and first lines can be combined to obtain
\begin{align}
   \bigg\{\bar\nabla^\mu \Big(H_{i_0i}\partial_\mu\Phi^i\Big)&-\frac{1}{2}\partial_{i_0}H_{ij}\partial_\mu \Phi^i\partial^\mu\Phi^j\bigg\}\bigg(\partial_\nu \Phi^{i_0}+L_{i_0}^{-1}L_\hi\partial_\nu \Phi^\hi -2V_{i_0}^{-1}\partial_\nu (L_{i_0}^{-1}V_{i_0})\bigg)\nonumber \\
   &-2 L_{i_0}^{-1} V_{i_0}  \bar\nabla_\nu \bigg\{ V_{i_0}^{-1}\Big( \bar\nabla_\mu(H_{i_0i} \partial^\mu \Phi^i)-\frac{1}{2}\partial_{i_0} H_{ij} \partial_\mu\Phi^i\partial^\mu\Phi^j\Big)\bigg\}=0.\label{tt}
\end{align}
Hence, choosing $L_{i_0}(\Phi)$ such that 
\be
\partial_\nu \Phi^{i_0}+L_{i_0}^{-1}L_\hi\partial_\nu \Phi^\hi -2V_{i_0}^{-1}\partial_\nu (L_{i_0}^{-1}V_{i_0})=0, \label{kdv2}
\ee
\Eq{tt} reduces to 
\begin{align}
 \bar\nabla_\nu \bigg\{ V_{i_0}^{-1}\Big( \bar\nabla_\mu(H_{i_0i} \partial^\mu \Phi^i)-\frac{1}{2}\partial_{i_0} H_{ij} \partial_\mu\Phi^i\partial^\mu\Phi^j\Big)\bigg\}=0,
\end{align}
which yields upon integration
\begin{align}
 \bar\nabla_\mu(H_{i_0i} \partial^\mu \Phi^i)-\frac{1}{2}\partial_{i_0} H_{ij} \partial_\mu\Phi^i\partial^\mu\Phi^j=\nu_{i_0}V_{i_0},
\label{V02}
\end{align}
where $\nu_{i_0}$ is a constant. For this result to coincide with \Eq{eq022} for $k=i_0$, we take 
\begin{equation}
    V_{i_0}=-\frac{1}{2\nu_{i_0}} \partial_{i_0}\bar U. 
\end{equation}

In that case, \Eq{V02} implies
\be
-2L_{i_0}^{-1}\Big(2 \bar\nabla_\mu(H_{i_0i} \partial^\mu \Phi^i)-\partial_{i_0} H_{ij} \partial_\mu\Phi^i\partial^\mu\Phi^j\Big)=2L_{i_0}^{-1}\partial_{i_0}\bar U, 
\ee
whose r.h.s. should equal  $\bar U$, for \Eq{eq0112} to match with the metric equation of motion~(\ref{eq012}). As a result, we necessarily have 
\be
L_{i_0}=2\, {\partial_{i_0}\bar U\over \bar U}.
\ee
Since $i_0$ is chosen arbitrarily, we conclude that
\be
L_{i}=2\, {\partial_{i}\bar U\over \bar U}
\ee
is required for all scalar equations of motion~(\ref{eq022}) to be satisfied, where $i\in\{0,\dots,N\}$.

To conclude that we succeeded in reproducing all equations of motions~(\ref{eq012}) and~(\ref{eq022}), we still have to check if \Eq{kdv2} and its counterparts where $i_0$ is replaced by any $k$ are satisfied. In fact, using the expressions of  $V_{i_0}$ and $L_{i}$, \Eq{kdv2} becomes
\be
\partial_{i}\bar U\partial_\nu\Phi^{i}=\partial_\nu\bar U,
\ee
which is independent of $i_0$ and turns out to be true.

As a result, \Eq{cc12} becomes 
\begin{equation}
{ \partial_i\bar U\over \bar U}\,   \delta\Phi^i=  {1\over 2}\, \bar g_{\mu\nu}\delta \bar g^{\mu\nu}\quad \Longleftrightarrow\quad \delta \Big[\bar U(\Phi)\, \sqrt{-\bar g}\Big]\!=0,
\end{equation}
which yields in string-fame  
\be
\delta \Big[U(\Phi)\, \sqrt{-g}\Big]\!=0.
\ee


\section{Spectrum}
\label{spec}

Since in our approach a field combination is always non-dynamical, one is tempted to question if the number of propagating degrees of freedom is reduced or not. In the following, we answer this question by explicitly showing that imposing $\sqrt{-g}e^{-2\Phi}$ to be non-dynamical does not eliminate the dilaton degree of freedom appearing in the action
\be
    \widehat \S=\frac{1}{2 \kappa_{0}^{2}} \int \mathrm{d}^{D} x \sqrt{-g} e^{-2 \Phi}\Big(R+4 \partial_{\mu} \Phi \partial^{\mu} \Phi\Big).
    \label{ac-d}
\ee
Indeed, a massless graviton and a scalar degree of freedom are still propagating. 

For definiteness, let us consider a flat Minkowski vacuum with a constant dilaton, \ie
\begin{equation}
g_{\mu \nu}=\eta_{\mu \nu}, \qquad \Phi=\Phi_{0}={\rm constant}.
\label{min}
\end{equation}
We can expand the action (\ref{ac-d}) to second order in the fluctuations about this background as 
\be
  \widehat  \S\left[\eta+\delta g, \Phi_{0}+\delta \Phi\right]=\widehat\S\left[\eta, \Phi_{0}\right]+\frac{1}{2} \delta^{2} \widehat\S\left[\eta, \Phi_{0}\right]+\cdots\, .
\ee
The second-order variation is given by 
\be
    \delta^{2} \widehat\S\left[\eta, \Phi_{0}\right]=\frac{1}{2 \kappa_{0}^{2}} \int \mathrm{d}^{D} x \,e^{-2 \Phi_{0}}\left\{h^{\mu \nu}\left(\partial_{\mu} \partial^{\rho} h_{\nu \rho}-\frac{1}{2} \square h_{\mu \nu}\right)\right\}, ~\quad\where\quad~ h_{\mu\nu}=\delta g_{\mu\nu}.
\ee
 To obtain this result, we have implemented the condition ${1\over 2}g_{\mu \nu} \delta g^{\mu \nu}+2 \delta \Phi=0$ and set 
\be
    \delta \Phi=\frac{1}{4} h, \qquad h=h_{\mu}^{\mu}.
\ee
Using the standard decomposition of a symmetric tensor 
\be
    h_{\mu \nu}=h_{\mu \nu}^{T T}+\partial_{\mu} A_{\nu}^{T}+\partial_{\nu} A_{\mu}^{T}+\left(\partial_{\mu} \partial_{\nu}-\frac{1}{D} \eta_{\mu\nu} \square\right) B+\frac{1}{D} \eta_{\mu \nu} h,
\ee
where the fields $h_{\mu \nu}^{T T}$ and $A_{\mu}^{T}$ are respectively transverse-traceless and transverse,
\be
   h_{\mu}^{T T}=\partial^{\nu} h_{\nu \mu}^{T T}=0, \qquad \partial^{\mu} A_{\mu}^{T}=0 ,
\ee
we find that the second-order variation becomes
\begin{align}
\delta^{2} \widehat\S\left[\eta, \Phi_{0}\right]&=  \frac{1}{2 \kappa_{0}^{2}} \int \mathrm{d}^{D} x \,e^{-2 \Phi_{0}}\left\{-h^{T T \mu \nu} \square h_{\mu \nu}^{T T}+\frac{(D-1)(D-2)}{D^{2}} \,\Psi \square \Psi\right. \\
& \hspace{3.55cm}\left.+4 \,\frac{D-1}{D^{2}} \,h \square \Psi-\frac{D-2}{D^{2}}\, h \square h\right\} \\
&=  \frac{1}{2 \kappa_{0}^{2}} \int \mathrm{d}^{D} x e^{-2 \Phi_{0}}\left\{-h^{T T \mu \nu} \square h_{\mu \nu}^{T T}-S \square S+\frac{D-1}{D-2}\, \Psi \square \Psi\right\},
\end{align}
where
\be
\Psi=\square B, \qquad  S=\frac{\sqrt{D-2}}{D}\left(h-2\, \frac{D-1}{D-2} \Psi\right) .  
\ee
Therefore, the spectrum comprises a helicity-2 state $h_{\mu \nu}^{T T}$, a spin-0 state $S$ and a ghost state $\Psi$. The latter corresponds to the usual conformal ghost mode of general relativity, which here manifests itself as the longitudinal part of the spin-1 part of the metric fluctuations $h_{\mu \nu}$.


\section{Discussion and conclusion}
\label{cl}

Consistency of a string model at tree level requires the vanishing of the sigma-model beta-functions on the worldsheet. We have seen that these equations for the metric, the antisymmetric tensor and the dilaton can be recovered by extremizing a universal effective action. To achieve this, we need to impose that the field combination $\sqrt{-g}e^{-2\Phi}$ is non-dynamical. Therefore, the model-dependent central-charge deficit term $\sqrt{-g}e^{-2\Phi}\delta c$ in the Lagrangian no longer contributes and can be omitted. In fact, the central-charge deficit arises as an integration constant, when solving the equations of motion. On may think that a drawback of this approach is that all values of the integration constant are allowed, regardless of the existence or not of a string-theory model. For instance, in the description of the Witten black hole of \Sect{WBH}, the Kac-Moody level $k=8f_0$ should be integer. However, at the two-derivative level, only large values of $k$ should be considered, spanning effectively a continuum. 

 At tree level, the string models admit moduli fields parametrizing moduli spaces. To incorporate them in the effective description, we must add to the Lagrangian a sigma-model kinetic term that involves the moduli-space metric. Since the latter is model dependent (in particular, in models that are not maximally supersymmetric at tree level), the effective action is no longer universal, and the moduli-space metric has to be determined case by case at the string level. Moreover, at higher genus, the moduli-space metric admits quantum corrections and an effective potential $U$ can be  generated. In the alternative approach we outline in this work, the latter is not included in the Lagrangian. Instead, it appears in the combination of fields constrained to be non-dynamical, namely $\sqrt{-g}\, U$, which eventually leads to the appearance of the gradient of $U$ 
 in the scalar equations of motion. The moduli-space metric and effective potential must therefore be predetermined at the string level in order to obtain the field equations and the solutions. However, the key point is that once the dynamics stabilizes all moduli fields at a minimum of the effective potential, the sigma-model kinetic term vanishes and the potential $U$ reduces to a cosmological term that in this alternative description appears as an integration constant, as in ~\cite{wein}. 

We emphasize that the fact that $\sqrt{-g}\, U$ is non-dynamical means that its variation vanishes ($\delta[\sqrt{-g}\, U]=0$) when one extremizes the action to obtain the independent equations of motion. This condition respects the full diffeomorphism invariance, and so does our effective description. However, as in general relativity, nothing prevents us to use a particular frame such that $\sqrt{-g}\, U=\varphi(x)$, where $\varphi(x)$ is a fixed background. When we choose to work in such a frame, the residual gauge symmetry is the group of volume-preserving diffeomorphisms. Moreover, even though the combination $\sqrt{-g}\, U$ is non-dynamical, we have shown that the number of the propagating degrees of freedom is not reduced. 

Beyond tree level, our derivations do not rely on the fact that the Lagrangian and the non-dynamical field combination admit an expansion in powers of the string coupling $e^\Phi$. Therefore, all conclusions should remain valid at the non-perturbative level. On the other hand, our analysis was based on beta functions derived at leading order in $\alpha'$, thus  at the two derivative level. As part of future work, it would be interesting to see how higher-derivative terms in the effective action may be incorporated.    

We have shown that besides the standard approach for describing the low energy physics of string theory, there exists an alternative effective field theory description, without any effective potential in the action but with a non-dynamical field combination instead. Since both approaches yield the same field equations for all the degrees of freedom, it would be interesting to know if other criteria can distinguish them. Such criteria may be experimental or theoretical, at the level of string theory. 
In fact, the question already arises when one wants to decide whether the general theory of relativity or a version of unimodular gravity is preferable for describing the real world~\cite{wein,garriga}. In particular, even if general relativity and unimodular gravity are equivalent at the classical level, they yield distinct Wheeler-DeWitt equations at the quantum level~\cite{Unruh,Page} and thus different predictions concerning the wavefunction of the universe. 

Another question concerns the dimensional reduction of the unimodular-like string effective action. In certain cases (\eg toroidal compactifications without fluxes), the minimal unimodular description for the compactified theory, where the effective potential does not appear in the action but in a non-dynamical field combination instead, can be directly obtained \via the dimensional reduction of the corresponding unimodular description in higher dimensions. This is also true for the precise form of the non-dynamical field combination involving the determinant of the metric and the scalar fields: The non-dynamical field combination for the compactified theory can be directly obtained from the non-dynamical field combination in the ``parent'' theory. However, this direct relationship between the two minimal unimodular descriptions will not hold in the generic cases. In particular, turning on non-trivial fluxes associated with the antisymmetric RR fields may lead to the appearance of a potential for the scalar fields in the action for the compactified theory. So it seems that the passage to the minimal unimodular description, with no potential in the action, will have to be implemented at a new step. 

Let us however emphasize that the situation is similar in the conventional description. For us the starting point is the string model and the sigma model action associated with the compactification manifold. In the compactified theory, there are two kinds of fields. Those that can be seen as components of the higher dimensional fields and those that emerge at special points of the moduli space, and are associated with stringy states and wrapped branes becoming light. For example, after compactifying a ten-dimensional theory on a Calabi-Yau manifold, the metric decomposes in an obvious notation as $(g_{\mu\nu},g_{i\bar \jmath},g_{ij})$\footnote{Gauge fields of the form $g_{\mu i}, g_{\mu \bar \imath}$ also appear.} and therefore, we will have a four-dimensional graviton $g_{\mu\nu}$, $h_{(2,1)}$ complex scalars and $h_{(1,1)}$ real scalars. In addition, extra states may emerge at special points of the moduli space, which do not exist in the original higher dimensional effective action. 
The first class of fields are obviously accommodated in the unimodular restricted string effective action (after all, consistent dimensional reduction requires solving the higher dimensional equations, here the $\beta$-function equations). On the other hand, the new fields, as well as fluxes associated with compact cycles, which arise after compactification and do not appear in the effective action for the uncompactified theory, cannot be directly incorporated \via a dimensional reduction of the effective action. One must resort to the full string model and the moduli space to obtain the effective description. This was also the case in the conventional description, where extra fields must be incorporated in the action to account for new stringy states becoming light. A full analysis concerning the inclusion of such fields, as well as fluxes in our new setting, would require a full in depth analysis that goes beyond the scope of the present work.

Finally, we may ask whether the effective field theory with non-dynamical field combination we consider can be embedded in a general framework similar to supergravity \cite{Nagy1,Nagy2}. In other words, are such constraints compatible with local supersymmetry? This is a non-trivial question since the fact that $\sqrt{-g}\, U$ is non-dynamical should have consequences for the superpartners of the graviton and scalar fields.

\section*{Acknowledgements} 

A.K. and H.P. would like to thank the University of Cyprus for hospitality, while A.K and N.T. acknowledge hospitality by the Ecole Polytechnique where early stages of this work have been done. The work of A.K. is partially supported by the NTUA basic research grant no 65228100. The work of H.P. and N.T. is partially supported by the Cyprus Research and Innovation Foundation grant EXCELLENCE/0421/0362.

\end{document}